\documentclass[final]{cvpr}

\usepackage{times}
\usepackage{epsfig}
\usepackage{graphicx}
\usepackage{amsmath}
\usepackage{amssymb}
\usepackage{subfigure}
\usepackage{overpic}

\usepackage{bm}
\usepackage{algorithm}

\usepackage{algorithmicx}
\usepackage{algpseudocode}

\usepackage{enumitem}
\setenumerate[1]{itemsep=0pt,partopsep=0pt,parsep=\parskip,topsep=5pt}
\setitemize[1]{itemsep=0pt,partopsep=0pt,parsep=\parskip,topsep=5pt}
\setdescription{itemsep=0pt,partopsep=0pt,parsep=\parskip,topsep=5pt}

\usepackage[pagebackref=true,breaklinks=true,colorlinks,bookmarks=false]{hyperref}

\def\cvprPaperID{159} 
\def\confYear{CVPR 2020}

\graphicspath{{figures/}}

\begin{document}

\title{Variational Hierarchical Directed Bounding Box Construction for Solid Mesh Models}

\author{Rui Wang ~~~~~ Wei Hua ~~~~~ Gaofeng Xu ~~~~~ Yuchi Huo ~~~~~ Hujun Bao
}


\maketitle

\begin{abstract}
Object oriented bounding box tree (OBB-Tree for short) has many applications in collision detection, real-time rendering, etc. It has a wide range of applications. The construction of the hierarchical directed bounding box of the solid mesh model is studied, and a new optimization solution method is proposed. But this part of the external space volume that does not belong to the solid mesh model is used as the error, and an error calculation method based on hardware acceleration is given. Secondly, the hierarchical bounding box construction problem is transformed into a variational approximation problem, and the optimal hierarchical directed bounding box is obtained by solving the global error minimum. In the optimization calculation, we propose that combining Lloyd clustering iteration  in the same layer and MultiGrid-like reciprocating iteration between layers. Compared with previous results, this method can generate aired original solid mesh models are more tightly packed with hierarchical directed bounding box approximation. In the practical application of collision detection, the results constructed using this method can reduce the computational time of collision detection and improve detection efficiency.

\end{abstract}

\section{Introduction}
Hierarchical bounding volumes are used to approximate complex geometric objects with several bounding volumes with slightly larger volumes and simple shapes. In many occasions that require real-time computing (such as collision detection, real-time rendering, etc.) To replace the original geometric object to participate in the calculation, to achieve the purpose of simplifying the calculation. Taking collision detection as an example, in the process of traversing the bounding volume hierarchy tree, the basic geometry that is obviously impossible to intersect can be excluded early through the fast intersection test between bounding volumes. object, thereby effectively improving the speed of collision detection.

Due to the wide range of applications, people have carried out a lot of research on the construction of hierarchical bounding volumes. According to different application requirements, geometric bounding volumes of different basic shapes can be selected to construct the hierarchical optimization approximation of the scene. These geometric bounding volumes include: Sphere \cite{r1,r2} (sphere), Axial Bounding Box (AABB) \cite{r3}, Oriented Bounding Box (OBB) \cite{r4} and Oriented Polyhedrons (k-DOPs) \cite{r5} and so on.

When people choose the basic shape of the bounding volume, they often need to balance the tightness of the original geometric object with the simplicity of the bounding volume test. Bounding volumes with complex shapes, such as directed polyhedra (k-DOPs), can be achieved via tighter wrapping. However, more complex intersection calculation is required during testing; for simple bounding volumes, although the calculation is simple during testing, the wrapping effect of the original geometric object is poor, which reduces the test efficiency to a certain extent. Weghorst et al. \cite{r6} proposed an expression for evaluating the quality of hierarchical bounding volumes for accelerated ray tracing algorithms, and it is widely used in situations where hierarchical bounding volumes are used, including collision detection \cite{r4}. To measure a static evaluation function for the effectiveness of the hierarchical bounding volume, which is defined as

\begin{equation}
    e=n_v\cdot c_v + n_p \cdot c_p
\end{equation}
where $n_v$ is the number of bounding volume intersection calculations, $c_v$ is the cost of performing a bounding volume intersection calculation, $n_p$ is the number of patch intersection calculations, and $c_p$ is the cost of the intersection calculations.

For different applications, a variety of hierarchical bounding volumes with different basic shapes have been proposed. In the application of collision detection, the hierarchical directed bounding box proposed by Gottschalk \cite{r4} is one of the most commonly used hierarchical bounding volumes. Therefore, this paper Select the directed bounding box to construct the hierarchical bounding volume, and study the construction algorithm of the optimal hierarchical directed bounding box. For the bounding box, $c_v$ and $c_p$ in formula can be regarded as constants, so the cost of the evaluation function e is minimized That is to find a hierarchical bounding box with as small $n_v$, $n_p$ as possible.

For the number of times $n_p$ of patch intersections, if the hierarchical directed bounding box is closer to the original model in appearance and shape, then the probability of misjudgment in the hierarchical directed bounding box test is smaller, and finally it involves the fine intersection of the original model. The number of patches that can be accurately determined is also less. Based on this, we propose to use the outer volume as an error measure to measure the approximation degree between the hierarchical directed bounding box and the original model. The outer volume refers to the space that belongs to the bounding volume but does not belong to the original grid The spatial volume of the model. This error definition does not consider the overlap of the bounding box inside the object, and only considers the part of the bounding box outside the original model, which allows us to minimize the error to obtain the optimal approximation of the original mesh model, Hierarchical directed bounding Box. This paper gives the definition and calculation method of the outer volume and transforms the construction of the hierarchical directed bounding box into a variational approximation problem.

To reduce the number of intersections $n_v$ of bounding boxes, it is necessary to wrap the upper bounding box more tightly. This is because the tighter wrapping of the upper bounding box can be more Quickly remove those lower bounding boxes that do not need to participate in testing. Traditionally, people generally use top-down or bottom-up division or greedy merging
method to construct a hierarchical directed bounding box, due to the lack of effective metrics in the construction to control the level of the bounding box and its tightness to the original model approximation. Therefore, it is difficult to ensure that the final hierarchical directed bounding box is optimal. Based on the external volume metric proposed in this paper, we adopt a method similar to MultiGrid \cite{r7}
The multi-layer reciprocating iterative optimization of the bounding box between layers and the Lloyd \cite{r8} clustering iterative optimization method of the same layer of the bounding box, to achieve the smallest error in the global sense. By optimizing the solution, the hierarchical directed bounding box that wraps the original solid mesh model most closely can be obtained.

The novelty of this paper is that: firstly, by defining the error metric of the outer volume, the problem of constructing a hierarchical directed bounding box is formalized as a minimization
The optimization approximation problem of the outer volume; secondly, in the optimization solution, a tighter directional bounding box is obtained through the optimization iteration between multiple layers and the clustering iteration between the same layers.

\section{Related Work}
Hierarchical bounding box is used for accelerating rendering-based applications \cite{wang2013gpu,wangimplementation,huo2015matrix,huo2016adaptive,huo2020spherical,cho2021weakly,fan2021real,huo2021survey,huo2020adaptive,huo2022extension,kim2020single,li2021multi,an2021hypergraph,park2021meshchain,zhang2021powernet,li2020automatic}. The initial method of constructing a hierarchical bounding volume is to perform octree division of a mesh model in space \cite{r9}. When the division reaches a certain depth, it stops, and each non-empty leaf node containing a patch forms a bounding volume to form a bounding volume. Obtain the approximation of the original model. Obviously, the approximation effect of this bounding volume uniformly divided in the world coordinate system space is very poor. Subsequently, people proposed a method of constructing a hierarchical bounding volume with an object as the approximation object, which is referred to as a hierarchical bounding volume for short. Bounding volume hierarchies algorithm. Due to the relative simplicity of spherical parameters, there are many algorithms that use spheres to construct scene-level bounding volumes \cite{r1,r2}, but it is often difficult to obtain tight wrapping results. Based on the axial bounding box (AABB), the hierarchical bounding volume algorithm \cite{r3,r10} is very fast, but the approximation effect of the original model is not very good. The directed bounding box (OBB) can approximate the object more closely than the axial bounding box and the bounding sphere, and it is significantly However, the intersection detection between directed bounding boxes is more time-consuming than the intersection detection between axial bounding boxes or bounding spheres. Directed polyhedrons (k-DOPs) \cite{r5} can be regarded as having Further generalization to the bounding box, although it can enclose the original object more tightly than other bounding volumes, the intersection detection between directed polyhedra is obviously more complicated.

This paper mainly discusses the construction methods of directed hierarchical bounding boxes. In the construction of hierarchical directed bounding boxes, most of the existing methods use top-down subdivision or bottom-up merging strategies to build hierarchical structures. Top-down methods based on subdivision strategies do not consider whether the original model is actually approximated, but use such as the longest axis of the enclosed model \cite{r9} or the principal component analysis of the spatial position of the enclosed model. The bottom-up merging strategy can only optimize and analyze the local approximation situation when merging. Therefore, this kind of algorithm basically adopts the greedy algorithm, and the obtained approximation. The results are mostly local optima.

Cohen-Steiner et al. \cite{r11} proposed a variational patch approximation method for mesh approximation research, which transformed the solution of patch approximation into finding a variational energy optimum, and achieved good results. Wu et al. \cite{r12} inherited the Cohen-Steiner method and proposed a variational approximation method for approximating the mesh model with a high-order polynomial patch \cite{r2}, and Lu et al. \cite{r13} further extended it to the application of bounding ellipsoids. The method in this paper adopts the hierarchical optimization approximation based on the directed bounding box. The directed bounding box has more advantages than bounding spheres or ellipsoids. For more complex shapes, the analytic form in the literature \cite{r2} cannot be used for the calculation of the directed bounding box, so new methods need to be studied. In addition, the above methods for generating bounding volumes do not consider the construction of the hierarchical relationship between bounding volumes. This paper It is the first time to introduce variational methods into the construction of hierarchical bounding volumes.

\section{Hierarchical Directed Bounding Box Approximation}

In this paper, we use $X$ to denote a 3D mesh model, and $Y$ to denote a hierarchical directed bounding box. For a directed bounding box with a hierarchical structure, we use $O_{ij}$ to denote the $j$-th directed bounding box of the $i$-th layer in $Y$, use $Z_i={O_{i0},O_{i1},…,O_{in}}$ to represent the set of directed bounding boxes of the $i$-th layer in $Y$. In the case of only discussing the same layer, for convenience, we omit the subscript $i$ and use $Z= {O_j}$ represents the directed bounding box of the same layer. Since this paper studies the solid grid model, we use $\Omega(\cdot)$ to represent the volume of space occupied by the 3D object. Then a layer of the 3D grid model $X$ is directed The bounding box $Y$ can be defined as

Definition 1. $Y$ is called a hierarchical directed bounding box approximation of the triangular mesh model $X$, when $Y$ satisfies

\begin{equation}
    \forall x \in \Omega(X), x\in\Omega(Y),
\end{equation}

where $x$ is a three-dimensional point in space, and $Y=\bigcup_{i} Z_i=\bigcup_{i,j}O_{ij}$ is the union of a series of hierarchical directed bounding boxes.

In order to calculate the optimal hierarchical directed bounding box approximation, we define the part of the volume in the space that belongs to the hierarchical directed bounding box $Y$ but does not belong to the model $X$ as the error, and calls such a spatial volume the outside mesh volume, Represented by $OMV(X,Y)$,

\begin{equation}
    \begin{aligned}
    OMV(X,Y)=\int_{R^3}g(x,X,Y)dx\\
    g(x,X,Y)=1, \quad if \quad x \notin \Omega(X) \quad and \quad x \in \Omega(Y) \\
    g(x,X,Y)=0, \quad others \\
    \end{aligned}
\end{equation}

It is difficult to solve such an external volume accurately, quickly and efficiently for complex mesh models. We first introduce a single directed bounding box $O$. For the outer volume of the mesh model $X$, the outer volume of the directed bounding box belonging to the same layer in the hierarchy is further given, and finally the calculation layer is given. A method for calculating the outer volume of a sub-directed bounding box.

For the outside bounding object volume of a single directed bounding box $O$, we denote it by $OBV(X, O)$, which is similar to the calculation like formula (3):

\begin{equation}
    \begin{aligned}
    OMV(X,Y)=\int_{R^3}g(x,X,O)dx\\
    g(x,X,Y)=1, \quad if \quad x \notin \Omega(X) \quad and \quad x \in \Omega(O) \\
    g(x,X,Y)=0, \quad others \\
    \end{aligned}
\end{equation}

According to formula (4), the external volume of multiple directed bounding boxes $Z={O_0,O_1,…,O_n}$ belonging to the same layer to the original model X can be expressed as

\begin{equation}
    \begin{aligned}
    OMV(X,Z)=OMV(x,\bigcup_j O_j)\\
    =\sum_j OBV(X,O_j)-Overlap(OBV(X,\bigcup_j O_j)),\\
    \end{aligned}
\end{equation}

where $Overlap(OBV(X,\bigcup_j O_j))$ is the overlapping part of the bounding box outside the original model $X$. Note that the volume of the overlapping part exist. That is,  $Overlap(OBV(X,\bigcup_j O_j ))\geq 0$, so $\sum_{j}OBV( X, O_j ) \geq OMV(X, Z )$. Accordingly, $\sum_j OBV( X ,O_j )$ can be used as the exact outer volume $OMV(X,Z)$'s upper bound estimation. In optimization, reducing $\sum_j OBV(X,O_j)$ can also reduce the exact outer volume of $X$ and $Z$. The computation of $\sum_j OBV( X , O_j )$ is much easier than directly computing the exact outer volume. In practical calculations, we use the sum of the outer volumes of the bounding box in $Z$ and the exact outer volume in place of $Z$ to speed up the computation.

\begin{equation}
    OMV(X,\bigcup_j O_j) \approx \sum_j OBV(X,O_j).
\end{equation}

For the total outer volume of the hierarchical directed bounding box, the upper bounding box that is tightly packed can effectively reduce the calculation amount of the intersection test of those lower bounding boxes that do not need to be calculated. Therefore, the upper bounding box approximates the original model to a higher degree than the lower bounding box. The degree of approximation of the original model is more important. We define the total outer volume of the hierarchical directed bounding box by introducing weights related to the hierarchy:

\begin{equation}
    e=\sum_{i=1}^{n_v} \omega_i(i)\cdot OMV(X,Z_i),
\end{equation}
where $\omega_i(i)$ is the weight function related to the layer where $Z_i$ is located, and the bounding box of the upper layer has a larger weight.

To sum up, by introducing the external volume, we formulate the hierarchical directed bounding box construction problem of the triangular mesh model as the following variational optimization problem, 

\begin{equation}
    \min e = \sum_{i,j} \omega_j(i)\cdot OBV(X,O_{ij})=\sum_{i,j} \omega_i(i)\cdot\int_{R^3}g(x,X,O_{ij})dx.
\end{equation}

That is to find an optimal bounding box to approximate $Y$, so that the overall error of is minimized.

\section{Optimal Solution of Hierarchical Directed Bounding Box}
\subsection{Exterior Volume Calculation of Single Bounding Box}

\begin{figure*}
    \centering
    \includegraphics[width=0.9\textwidth]{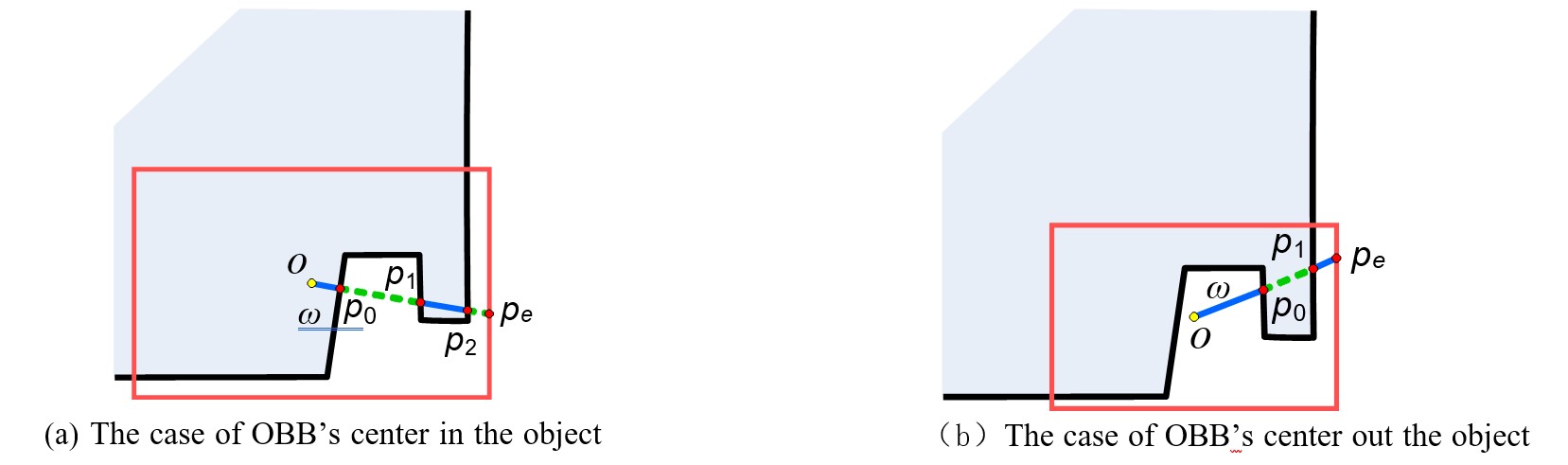}
    \caption{The computation of signed outside bounding object volume.}
    \label{fig1}
\end{figure*}

Due to the complexity of the grid model, we consider a discrete method to calculate formula (4). First, consider the case where the center of the bounding box is inside the original grid model $X$, as shown in Figure \ref{fig1}(a), we send out from the center of the bounding box to the direction $\omega$ an intersection ray $l$. The inteersction points with the mesh model $X$ are recorded as $p_0, p_1, ..., p_n$, respectively, and finally intersect with the boundary of the bounding box at the point $p_e$. Then the part $l_out$ of the line segment $op_e$ outside the mesh model $X$ can be calculated by the following:

\begin{equation}
    l_{out}=p_e o - \sum_i sign (n(p_i)\cdot \omega)p_i o
\end{equation}
where $sign$ is the sign function, and the contribution of each line segment to the external volume can be judged according to the relative direction of the normal at point $p_i$ and the ray $l$. For example, in Figure \ref{fig1}(a) the length of the line segment outside the grid (indicated by the dotted line) is $l_{out}=p_eo-(p_2o-p_1o+p_0o)$.

The total outer volume can be obtained by integrating $\omega$ in all directions in local coordinates centered on the bounding box:

\begin{equation}
OBV(X,O)=\int_\Omega l_{out}(\omega)d\omega.
\end{equation}

When the center of the bounding box is outside the mesh model, we use the same method as above to calculate the volume inside the mesh model (dashed line segment in Figure \ref{fig1}(b)). Finally, subtract the inner volume from the total volume of the bounding box to get the outer volume.

In practice, we use GPU to accelerate the calculation: take the center of the bounding box as the viewpoint, map the spatial direction to the CubeMap of $6×m×m$ (for example, $m=32$), one pixel on each surface of the CubeMap corresponds to A direction cone; using each face of the CubeMap as the projection plane, by drawing the bounding box $O$ and the grid model $X$, the volume of the direction cone inside and outside the grid is calculated in the shader; finally, by adding and subtracting the volume of these tetrahedrons Obtain the external volume. We adopt a 4-step calculation process: 1) Draw the triangular mesh $T_O$ of the bounding box, calculate the depth of the intersection $p_e$ of each pixel direction $l(\omega)$ and $T_O$, and calculate the direction cone corresponding to $p_eo$ according to the depth 2) Draw the triangular mesh $T_X$ of the original model, and send the depth texture of the bounding box in each discrete direction obtained in the first step into the drawing pipeline. Use the vertex shader and fragment shader to calculate the intersection of the light and $T_X$ The depth of $p_i$ and the normal direction at the intersection point, and compare the depth of $p_i$ with the depth of the intersection $p_e$ of the bounding box passed in in this direction in the fragment shader. If the depth of $p_i$ is less than the depth of $p_e$, calculate the corresponding depth in this direction. The volume of the direction cone. If the depth of $p_i$ is greater than the depth of $p_e$, it means that the intersection point is outside the bounding box and does not participate in the calculation. Using the Blend operation of Opengl, the volume of the direction cone calculated by all intersection points is summed and output as Texture. 3) The third step, align the textures of the first and second steps and draw, and use the fragment shader to do subtraction to get the final external volume in each direction, which is output as color. 4) It will represent the external volume in each direction The color results are read back from the graphics card and summed to get the total external volume.

\subsection{Optimal Solution of Hierarchical Directed Bounding Box Approximation}
\subsubsection{Optimal Solution of Bounding Box Approximation in Hierarchy}

\begin{figure*}
    \centering
    \includegraphics[width=0.6\textwidth]{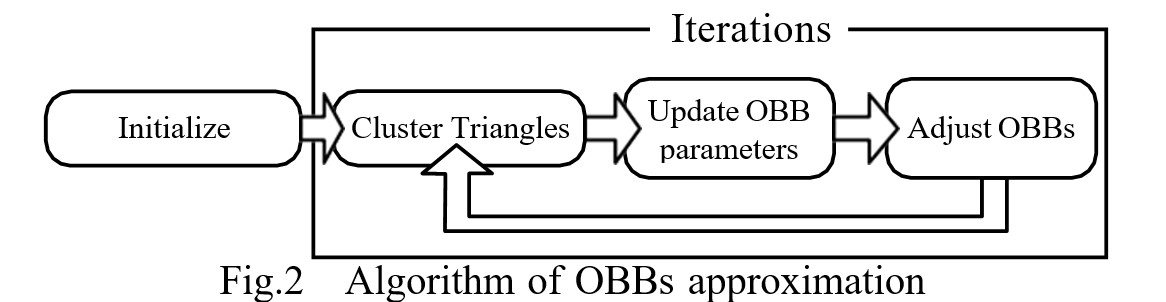}
    \caption{Algorithm of OBBs approximation.}
    \label{fig2}
\end{figure*}

Due to the complexity of the 3D grid model, the algorithm that directly solves a set of directed bounding boxes in 3D space to optimize the approximation to the grid model has high complexity and is difficult to calculate. Cohen-Steiner \cite{r11} proposed new approximation approach, which transforms the patch approximation problem into a global optimization of patch partitioning. Based on their work, we propose a variational directed bounding box for mesh models Approximation algorithm.

For all patches $F$ of the triangular mesh model $X$, find a division of $F$, ${F_0,F_1,…,F_N}$, satisfying $F=\bigcup_i F_i$ and $F_i\bigcap F_j=\emptyset$, $i \neq j$, such that the outer volume $\sum_i OBV(X ,O_i)$ is minimized when the bounding box $O_i$ surrounds $F_i$. That is

\begin{equation}
    \begin{aligned}
    \arg\min_{F_0,F_1,...,F_n} e(X,Z)=\sum_j OBV(X,O_j)\\
    s.t. \quad F=\bigcup_i F_i; F_i\cap F_j=\emptyset,i\neq j; F_i\in O_i\\
    \end{aligned}
\end{equation}

For the partitioning this formular, we use the iterative Lloyd \cite{r8} clustering algorithm to solve the problem. The flow of the algorithm is shown in Figure \ref{fig2}. Since the underlying constraint of the directed bounding box is the need to conservatively enclose the original model, when After the spatial position and the directed axis of the directed bounding box are determined, the shape parameters of the bounding box are also uniquely determined for the determined patch division. Therefore, in the calculation, we only need to optimize the position of the directed bounding box and the directed bounding box with 6 parameters to the axis.

\begin{figure*}
    \centering
    \includegraphics[width=0.9\textwidth]{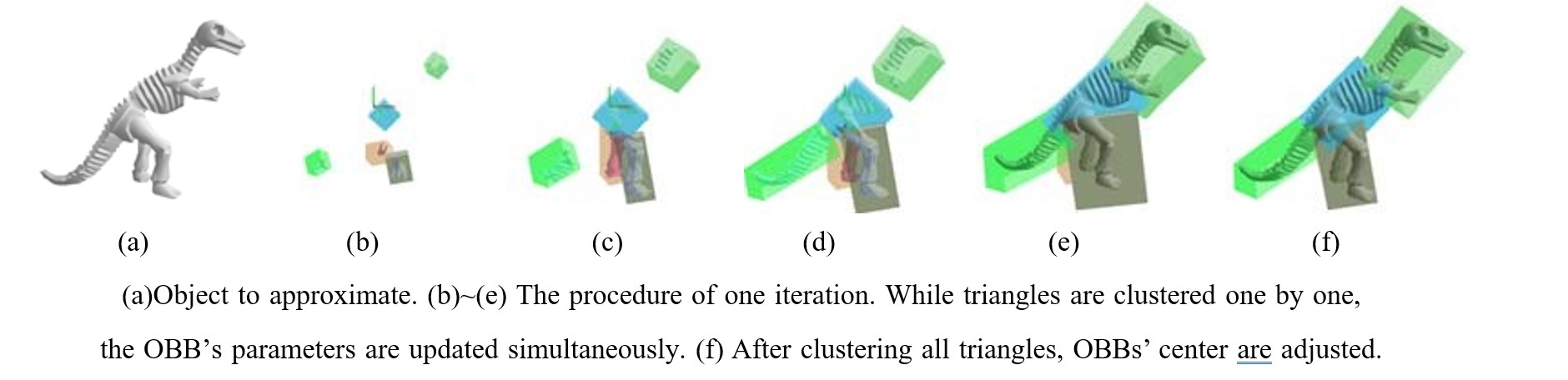}
    \caption{One iteration step in OBBs approximation.}
    \label{fig3}
\end{figure*}

According to Lloyd iteration:

Initialization. For each bounding box, we randomly select one or several patches as seeds to start partitioning. Figure \ref{fig3}(b) shows the initial seeds.

Iterative optimization. Our iteration consists of 3 steps: compute a new partition based on the current bounding box parameters (center and directed axis), update the parameters of each bounding box based on the new partition, if necessary The partition is adjusted to avoid getting trapped in local minima. Figure \ref{fig3}(c~f) illustrates an iterative optimization process.

1) Patch division. During patch division, the center of the bounding box remains unchanged. We use a filling algorithm (Flooding): maintain a priority queue to save the distance from each patch to the nearest bounding box; at the beginning, for the center of each bounding box, we select a patch closest to it as a seed and push it into the queue; when processing, each time a patch with the smallest distance $t_{min}$ to the center of the nearest bounding box is taken from the queue, and its adjacent faces The patch is put into the queue; for the patch, we divide it into different bounding boxes $O_i$ respectively, and calculate the outer volume change $dOBV(X,O_i)$ of the bounding box caused by the change of the shape parameter of $O_i$ from $S_i$ to $S_i'$ after dividing it from $t_{min}$ to $O_i$.
\begin{equation}
    dOBV(X,O_i)=\|OBV(X,O_i(S_i')-OBV(X,O_i(S_i))\|.
\end{equation}

Our partitioning principle is to partition tmin to the bounding box $O_k$ with the smallest external volume change, that is, $O_k$ satisfies:
\begin{equation}
    \forall i\in{1,2,...,N}dOBV(X,O_i)\leq dOBV(X,O_i).
\end{equation}

\begin{figure*}
    \centering
    \includegraphics[width=0.9\textwidth]{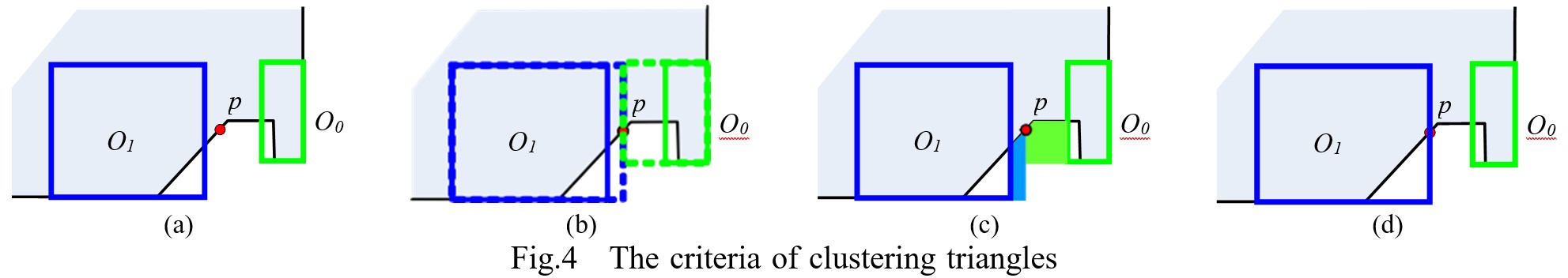}
    \caption{The criteria of clustering triangles.}
    \label{fig4}
\end{figure*}

Figure \ref{fig4} shows the criterion of patch division by a point division. The division of a patch does not depend on its distance from the nearby bounding box, but according to the minimum external volume change. When point $p$ has two potential bounding boxes, $O_0$ and $O_1$(Figure \ref{fig4}(a)), the outer volume caused by the division of $O_0$ and $O_1$ respectively (Figure \ref{fig4}(b), (c)) id divided to $O_1$ with the smallest volume increase (Figure \ref{fig4}(d)).

In the specific implementation, we maintain a list for each bounding box to record the calculated shape parameters and the corresponding external volume. In the actual calculation, this list can greatly simplify the calculation and accelerate the partitioning process when dividing the patch, and improve the approximate efficiency of optimization calculations.

2) Bounding box parameter update. When a new patch division is obtained, for each bounding box there is a patch set $F_i$, we need to find the optimal directed bounding box parameters to minimize the outer volume surrounding $F_i$. It is difficult to solve it analytically, so we use the numerical discrete solution method. The gradient direction of the point is obtained by perturbing the method near the initial parameters, and then the steepest descent method is used to solve the optimal bounding box parameters. Although the parameters of the directed bounding box The dimension is 6, but since we use a GPU-accelerated method to quickly calculate the outer volume under a certain parameter, updating the bounding box parameters using a discrete method can be done quickly.

3) Adjustment of bounding box. Due to the problem of initial value selection, the above iterative process sometimes gets stuck in a local minimum and cannot continue. When the total outer volume changes very little in several iterations, we sort the number of bounding patches, the size of the overlapping area between bounding boxes, etc.,  according to the importance of bounding boxes, and a new directed bounding box approximation is obtained by adjusting the least important bounding box. If the error is smaller than before the adjustment, then we continue the optimization iteration from the adjusted result, otherwise the adjustment is rejected and the optimization is exited. After adjusting the bounding box, our method can better bypass the local minimum in the optimization and obtain better results.

\begin{figure*}
    \centering
    \includegraphics[width=0.9\textwidth]{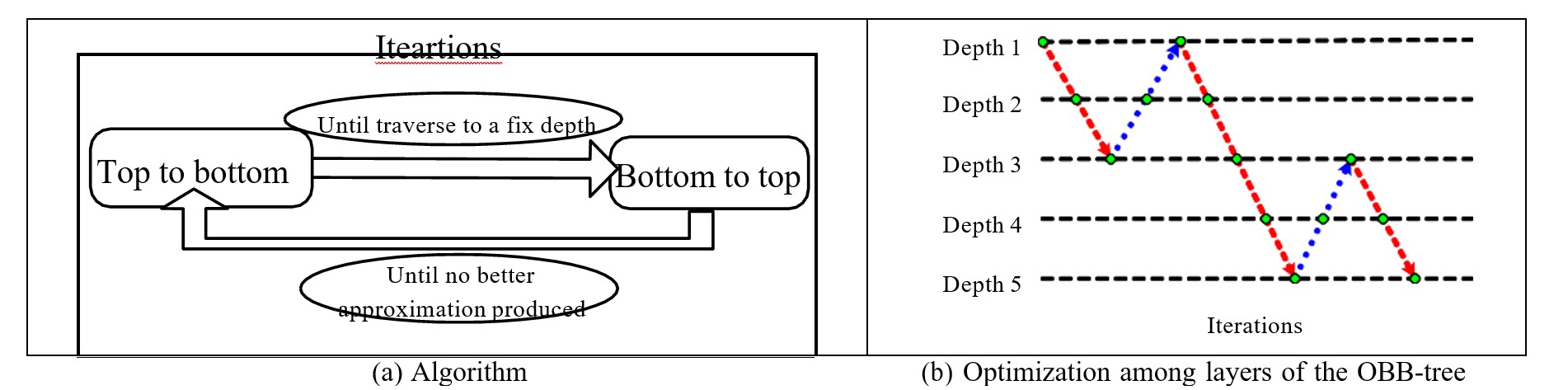}
    \caption{Optimization of hierarchical OBBs.}
    \label{fig5}
\end{figure*}

\subsubsection{Optimal Solution of Bounding Box Approximation between Levels}
Based on the calculation method of the directed bounding box at the same layer, and drawing on the idea of the MultiGrid \cite{r7} method for solving multivariable differential problems, we propose an iterative solution algorithm for multi-resolution reciprocating optimization for the approximation of hierarchical bounding boxes.

The algorithm flow is shown in Figure \ref{fig5}(a), and Figure \ref{fig5}(b) illustrates the optimization of the algorithm between different layers in the solution iteration. The dot is an iteration, and the direction of the short arrow indicates that the top-down optimization strategy is adopted. The dotted arrows indicate that the bottom-up optimization strategy is adopted. Through repeated optimization iterations between different levels, under the premise of ensuring the computational efficiency, the optimization solves the hierarchical bounding box approximation with the smallest error (formula (7)).

Top-down decomposition optimization strategy: the number of bifurcations of the hierarchical directed bounding box is given in advance. For a directed bounding box $O_i$ given a patch partition, we use the method in the previous subsection to solve its lower bounding box. One-level bounding box approximation. Top-down decomposition optimization recursively solves to a predetermined depth and then switches to bottom-up merge optimization.

Bottom-up merging optimization strategy: The object of merging optimization is the bounding box of the lower layer corresponding to a certain level node. During merging optimization, starting from the bounding box of the current level as the bounding box of the leaf node, the same layer as the previous section is adopted layer by layer. The method of solving the bounding box approximation is similar to the combined optimization calculation, but in the calculation, the patch is no longer used as the division unit, but the bounding box is used as the division unit. By calculating the optimal division of these bounding boxes, after each merging, we calculate the error of the merged hierarchy according to formula (7). If the result is not as good as before merging, then abandon the merging optimization and switch to the top-down decomposition optimization. Good before merging, then continue to merge and optimize the upper nodes.

Compared with the traditional optimization solution of hierarchical bounding boxes, three methods are generally used: bottom-up and step-by-step insertion of hierarchical nodes. Compared with the top-down method \cite{r5,r10,r14,r15}, the traditional methods are basically based on greed ( greedy) strategy, which only considers the current layer for optimization, and cannot guarantee the optimization of the overall error; our method can achieve the minimum overall error by defining the overall error and reciprocating iterative optimization between layers. However, since our method requires different layers Compared with the traditional method, it requires more computation time.

\section{Result and Comparison}

\begin{figure*}
    \centering
    \includegraphics[width=0.9\textwidth]{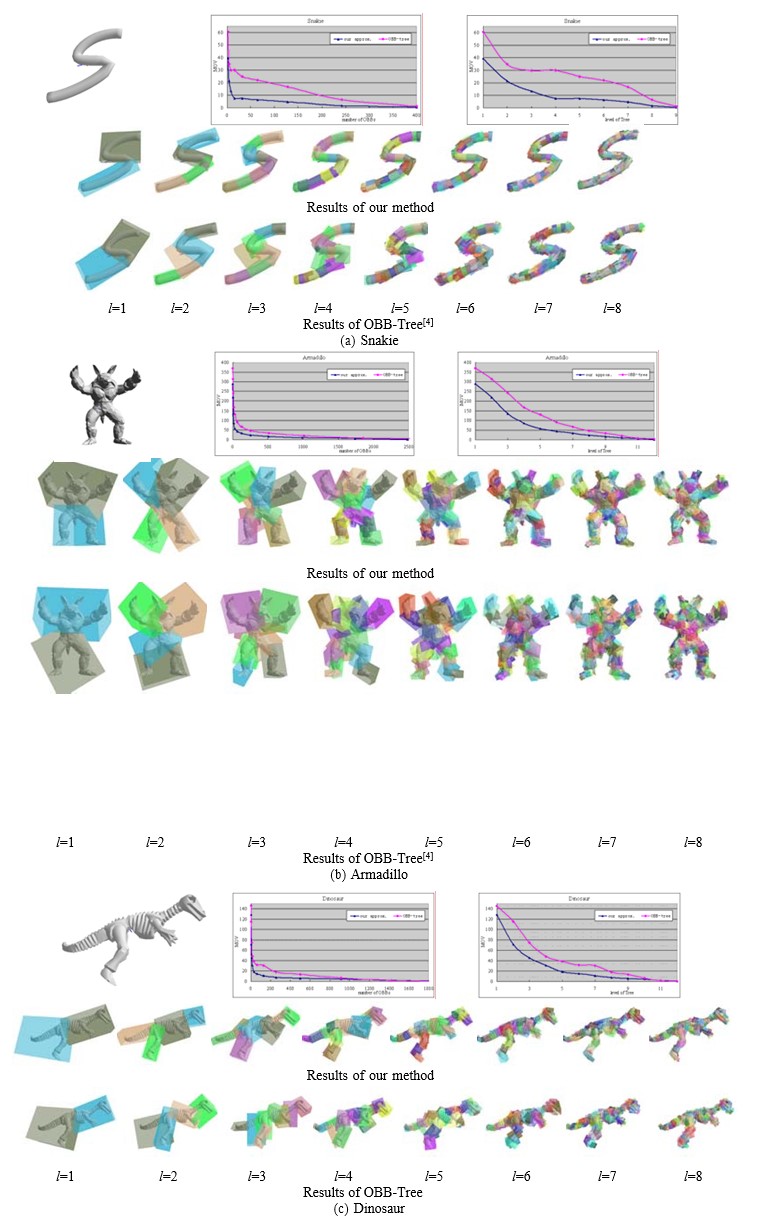}
    \caption{Results comparison of ours and OBB-Tree’s.}
    \label{fig6}
\end{figure*}

In the experiment, we choose to build a binary hierarchical directed bounding box. In the calculation, we take $\omega_i=l_{sub}$ as the weight between the layers, where $l_{sub}$ is the number of layers from the layer to the leaf node, for example, the leaf node’s $\omega_i=1.0$; $\omega_i=2.0$ of the node at the top of the leaf node, and so on. In the experiments later in this chapter, the above simple approach achieves good results.

Although we have adopted some acceleration methods, this method requires more computation time than the traditional directed hierarchical bounding box construction method. In our experiments, for 8k~10k models, the GPU takes 20~80 milliseconds to perform an outer volume calculation, but due to the large amount of external volume calculation required for patch division and bounding box parameter update, it takes 2~ 3 hours to construct a 12-layer binary directed hierarchical bounding box (the number of leaf nodes is $2^{12}$).

Since OBB-tree \cite{r4} is the most widely used directed hierarchical bounding box algorithm, and the author provides the source code RAPID (http://www.cs.unc.edu/~geom/OBB/OBBT.html), so our comparison is conducted around it. Comparisons are made from the appearance and geometric errors of the generated hierarchical bounding boxes and the computational efficiency in collision detection applications.

Figure \ref{fig6} shows the approximation results of multiple models using our method OBB-Tree method, showing the approximation results of the directed bounding box of the three models "Snakie", "Armadillo" and "Dinosaur" at different levels. We only show the results of bounding boxes for layers 1-8. From the results, our method is better in terms of the error comparison between the visual results and the outer volume.

\begin{figure*}
    \centering
    \includegraphics[width=0.6\textwidth]{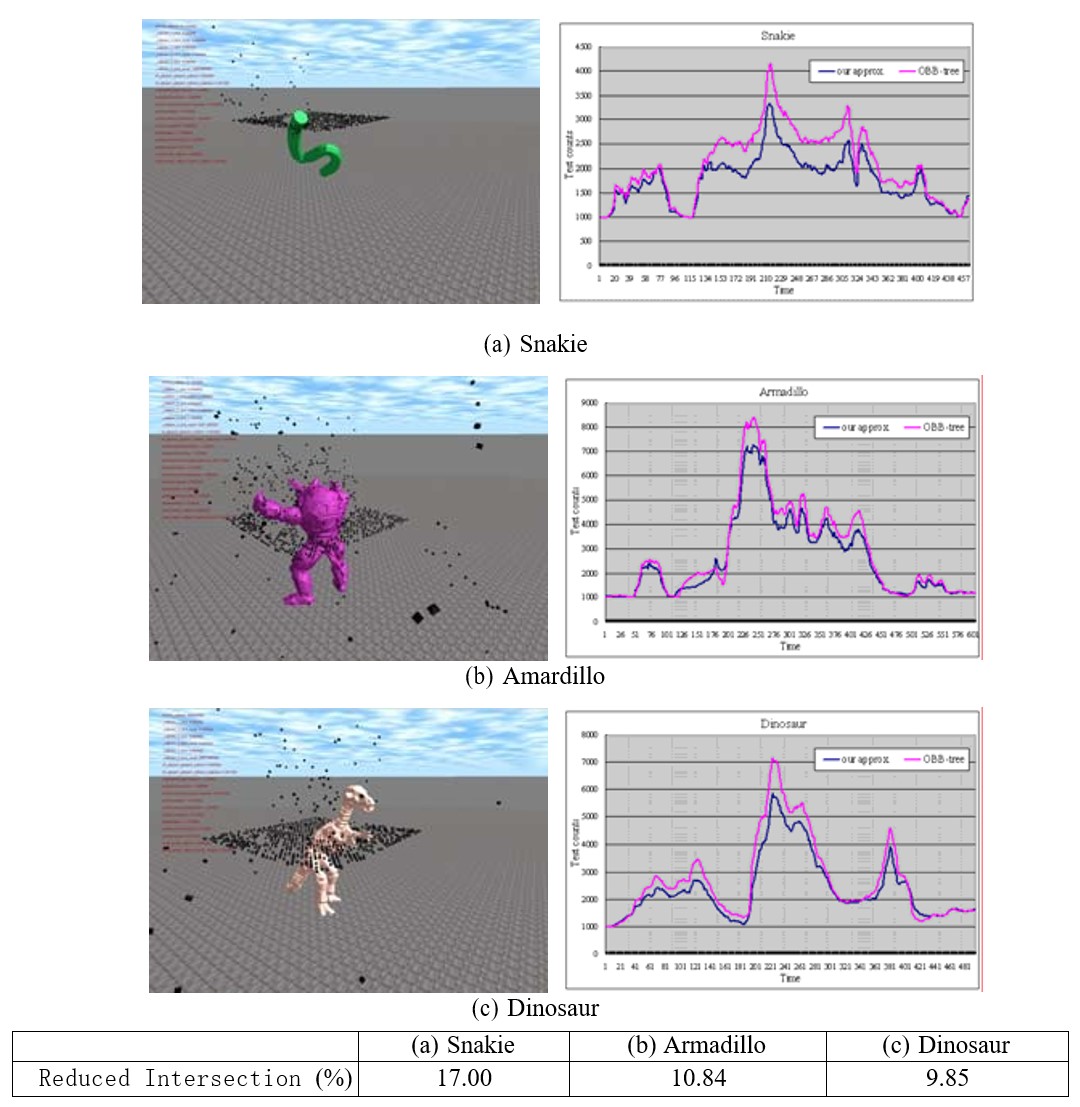}
    \caption{Performance comparison in collision detection between ours and OBB-Tree’s.}
    \label{fig7}
\end{figure*}

In order to detect the practical application of the approximation results in collision detection, we build a collision detection test environment (Figure \ref{fig7}) based on the open source physics engine library ODE (http://www.ode.org/). In this environment, the model revolves around Its own center rotates, and a collider falls from the top. We test the intersection between the collider and the model. If the model collides with the collider, the collider will be bounced. In the collision detection, we use our method to generate the The hierarchical bounding box and the hierarchical bounding box generated by the OBB-tree \cite{r4} method are used for rough calculation of intersection. Taking one collider-bounding box intersection as an intersection calculation unit, record the rough calculation of the hierarchical bounding box in unit time. The total number of collider-bounding box intersections required for the intersection. From the perspective of real-time performance, the smaller the number of intersections, the better the approximation effect of the hierarchical bounding box. In the comparison, we test a longer time, choose a large number (> 1000) colliders drop randomly. The results of the number of intersections/time are shown in Figure \ref{fig7}. As can be seen from the figure, under a large number of intersection tests, our hierarchical bounding box is better than the results generated by the OBB-tree \cite{r4} method. Even better, for the 3 models we tested, the number of intersections can be reduced by $10\%~17\%$.

\section{Conclusion}
This paper mainly discusses the hierarchical structure construction of the directed bounding box approximation of the solid mesh model. It is proposed to use the outer volume as a measure to measure the tightness of the bounding box approximation to the mesh model, thereby transforming the construction of the hierarchical bounding box into a variational approximation problem. In the optimization calculation of the optimal hierarchical bounding box approximation, we use the multi-layer reciprocating iteration similar to MultiGrid \cite{r7} and the Lloyd \cite{r8} clustering iterative method to solve the variational optimization with the smallest error in the global sense to obtain the optimal solution. The original solid mesh model is the most tightly packed hierarchical directed bounding box. Compared with the previous results, our method is better.

The biggest drawback of this method is that the calculation time is long. Compared with the traditional method, it is more suitable for occasions where the quality of the approximation results is higher, but the preprocessing time is less restricted. Although some acceleration methods have been considered in this paper, improving computational efficiency is still one of the focuses of the next research. In addition, this method uses the external volume as the error metric, so it can only deal with solid mesh models with spatial volumes, but in practical applications, a large number of meshes are not closed models, therefore, our next step will be to target the hierarchical directed bounding box construction method for a more general grid model and carry out the next step.

There are some extended articles about applying physical lighting computation in various applications:

\begin{enumerate}
    \item Deep Learning-Based Monte Carlo Noise Reduction By training a neural network denoiser through offline learning, it can filter noisy Monte Carlo rendering results into high-quality smooth output, greatly improving physics-based Availability of rendering techniques \cite{huo2021survey}, common research includes predicting a filtering kernel based on g-buffer \cite{bako2017kernel}, using GAN to generate more realistic filtering results \cite{xu2019adversarial}, and analyzing path space features Perform manifold contrastive learning to enhance the rendering effect of reflections \cite{cho2021weakly}, use weight sharing to quickly predict the rendering kernel to speed up reconstruction \cite{fan2021real}, filter and reconstruct high-dimensional incident radiation fields for unbiased reconstruction rendering guide \cite{huo2020adaptive}, etc.
    \item The multi-light rendering framework is an important rendering framework outside the path tracing algorithm. Its basic idea is to simplify the simulation of the complete light path illumination transmission after multiple refraction and reflection to calculate the direct illumination from many virtual light sources, and provide a unified Mathematical framework to speed up this operation \cite{dachsbacher2014scalable}, including how to efficiently process virtual point lights and geometric data in external memory \cite{wang2013gpu}, how to efficiently integrate virtual point lights using sparse matrices and compressed sensing \cite{huo2015matrix}, and how to handle virtual line light data in translucent media \cite{huo2016adaptive}, use spherical Gaussian virtual point lights to approximate indirect reflections on glossy surfaces \cite{huo2020spherical}, and more.
    \item Automatic optimization of rendering pipelines Apply high-quality rendering technology to real-time rendering applications by optimizing rendering pipelines. The research contents include automatic optimization based on quality and speed \cite{wang2014automatic}, automatic optimization for energy saving \cite{ wang2016real,zhang2021powernet}, LOD optimization for terrain data \cite{li2021multi}, automatic optimization and fitting of pipeline rendering signals \cite{li2020automatic}, anti-aliasing \cite{zhong2022morphological}, etc.
    \item Using physically-based process to guide the generation of data for single image reflection removal \cite{kim2020single}; propagating local image features in a hypergraph for image retreival \cite{an2021hypergraph}; managing 3D assets in a block chain-based distributed system \cite{park2021meshchain}.
\end{enumerate}

\bibliographystyle{ieee}
\bibliography{srbib}

\end{document}